\documentclass[twocolumn,showpacs,aps,prl,superscriptaddress]{revtex4}

\usepackage{graphicx}
\usepackage{dcolumn}
\usepackage{multirow}
\usepackage{amsmath}
\usepackage{epsfig} 

\input pubboard/babarsym

\newcommand{\BABARPubYear}    {06}
\newcommand{\BABARPubNumber}  {18}

\newcommand{\SLACPubNumber} {11855}

\newcommand {\mX}{\ensuremath{m_{X}}}

\newcommand {\bre}{\ensuremath{B_{\rm reco}}}

\newcommand {\Dsjsm}{\ensuremath{D_{sJ}(2460)^-}}
\newcommand {\Dsm}{\ensuremath{D_{s}^-}}
\newcommand {\Dssm}{\ensuremath{D_{s}^{*-}}}
\def\etal {{\it et al.}}

\def\figurebox#1#2#3{%
    \def\arg{#3}%
    \ifx\arg\empty
    {\hfill\vbox{\hsize#2\hrule\hbox to #2{\vrule\hfill\vbox to #1{\hsize#2\vfill}\vrule}\hrule}\hfill}%
    \else
    {\hfill\epsfbox{#3}\hfill}%
    \fi}

\begin{document}

\preprint{\babar-PUB-\BABARPubYear/\BABARPubNumber}
\preprint{SLAC-PUB-\SLACPubNumber}

\begin{flushleft}
\babar-PUB-\BABARPubYear/\BABARPubNumber\\
SLAC-PUB-\SLACPubNumber\\ 
\hepex{0605036}
\end{flushleft}

\title{
\vspace*{-1.0em}
{\boldmath \bf Study of $\B \ra D^{(*)}D_{s(J)}^{(*)}$ Decays and Measurement of \Dsm\ and \Dsjsm\ Branching Fractions} 
} 
%
\author{B.~Aubert}
\author{R.~Barate}
\author{M.~Bona}
\author{D.~Boutigny}
\author{F.~Couderc}
\author{Y.~Karyotakis}
\author{J.~P.~Lees}
\author{V.~Poireau}
\author{V.~Tisserand}
\author{A.~Zghiche}
\affiliation{Laboratoire de Physique des Particules, F-74941 Annecy-le-Vieux, France }
% ...
\collaboration{The \babar\ Collaboration}
\noaffiliation

\date{\today}
\begin{abstract}

We present branching fraction measurements of twelve $\B$ meson decays of the form $\B \ra D^{(*)}D_{s(J)}^{(*)}$.  
The results are based on  \Y4S decays in \BB pairs. One 
of the $B$ mesons is fully reconstructed and the other decays to two 
charm mesons, of which one is reconstructed, and the mass and momentum of the other is inferred by kinematics. 
Combining these results with 
previous exclusive branching fraction measurements, we determine 
$\BR(\Dsm\ra\phi\pim) = (4.62 \pm 0.36_{\rm stat.} \pm 0.51_{\rm syst.})\%$, $\BR(\Dsjsm\ra\Dssm\piz) = 
(56 \pm 13_{\rm stat.} \pm 9_{\rm syst.})\%$ and 
 $\BR(\Dsjsm\ra\Dsm\gamma) = (16 \pm 4_{\rm stat.}\pm 3_{\rm syst.})\%$. 
\end{abstract}
\maketitle
In this paper we present the study of charged and neutral $B$ mesons decaying to 
two charm mesons, i.e. $\Bb\to D_{\rm meas}D_X$~\cite{charge-conj}.
$D_{\rm meas}$ represents a fully reconstructed $D^{(*)+,0}$ or $D_{s}^{(*)-}$
meson, and the mass and momentum of the $D_X$ 
are inferred from the kinematics of the two-body $B$ decay. This study 
allows measurements of $B$ branching fractions without any assumption 
on the decays of the $D_X$.   Measurements of these two-body branching
fractions can provide tests of the factorization of the decay amplitudes\,\cite{Beneke} 
in the  high momentum transfer regime\;\cite{luo}.
From two separate classes of events with $D_{\rm meas}=D^{(*)-}_s$ and with $D_X=D^{(*)-}_s$
we measure the branching fraction of $\Dsm\ra\phi\pim$, which has important 
implications for a wide range of $D_s$ and \B physics.  Furthermore, 
we select final states with $D_X=\Dsjsm$ and combine with the \babar\ measurements of 
$\BR(\Bb \ra D^{(*)+,0}\Dsjsm) \times \BR(\Dsjsm\ra\Dssm\piz) $
and  $\BR(B \ra D^{(*)+,0}\Dsjsm) \times \BR(\Dsjsm\ra\Dsm\gamma) $~\cite{Aubert:2004pw}, 
thus extracting for the first time the absolute branching fractions of this recently observed 
state ~\cite{Besson:2003cp}.  

This analysis uses \upsbb events in which either a \Bp or a \Bz meson  
decays into a fully reconstructed hadronic final state (\bre ).
The measurements are based on an integrated luminosity of 
$210.5~\invfb$ recorded at the \Y4S resonance with the \babar\ detector at the \pep2\ asymmetric-energy
\epem\ collider operating near the \Y4S\ resonance.
An additional 21.7~\invfb\  recorded 40~\mev\ below the resonance ({\it off-resonance})
are used to evaluate backgrounds. 
The \babar\ detector is described in detail elsewhere~\cite{babarNIM}. 
Charged-particle trajectories are measured by a  vertex tracker with 5 double-sided layers  
and a 40-layer drift chamber, both operating in a 1.5-T magnetic field of a superconducting solenoid. 
Charged-particle identification is provided by the specific energy loss (\dedx) 
in the tracking devices and by an internally reflecting ring-imaging Cherenkov detector. 
Photons are detected by a CsI(Tl) electromagnetic calorimeter. 
We use Monte Carlo simulations (MC) of the \babar\ detector based on GEANT4~\cite{GEANT4} 
to optimize selection criteria and determine selection efficiencies.

To reconstruct a large sample of $B$ mesons, the hadronic decays $\bre
\rightarrow  \Db Y^{+}, \Db^* Y^{+}$  are  selected.  Here, the  system
$Y^{+}$ consists of hadrons with a total charge of $+1$, composed
of $n_1\pi^{\pm}\, n_2K^{\pm}\, n_3\KS\,  n_4\piz$, where $n_1 + n_2 \leq
5$,  $n_3  \leq  2$,  and  $n_4  \leq  2$.   We  reconstruct  $D^{*-}\ra
\Dzb\pi^-$; $\Dstarzb \ra
\Dzb\piz, \Dzb\gamma$; $D^-\ra K^+\pi^-\pi^-$, $K^+\pi^-\pi^-\piz$, $\KS\pi^-$,
$\KS\pi^-\piz$, $\KS\pi^-\pi^-\pi^+$; $\Dzb\ra K^+\pi^-$,
$K^+\pi^-\piz$, $K^+\pi^-\pi^-\pi^+$,  $\KS\pi^+\pi^-$; and  $\KS \ra \pi^+\pi^-$. 
The kinematic consistency of \bre\ candidates 
is checked with two variables,
the beam energy-substituted mass $\mes = \sqrt{s/4 -
\vec{p}^{\,2}_B}$ and the energy difference 
$\Delta E = E_B - \sqrt{s}/2$. Here $\sqrt{s}$ is the total
energy in the \FourS center-of-mass (CM) frame, and $\vec{p}_B$ and $E_B$
denote the momentum and energy of the \bre\ candidate in the same
frame.  The resolution on $\Delta E$ is measured to be $\sigma_{\Delta E}=10-35\mev$, depending on
the decay mode, and we require $|\Delta E|<3\sigma_{\Delta E}$.

For each reconstructed $B$ decay mode, the purity
${\cal P}$ is estimated as the ratio of the number of signal events with \mes$>
5.27$\gevc to the total number of events in the same range, 
and is evaluated on data.  We only use modes for which $\cal P$ exceeds a
decay-mode dependent threshold in the range of 9\% to 24\%. 
 In events with more than one
\bre\, we select the decay mode with the highest purity.  
On average, we reconstruct one signal $\bre$ candidate in 0.3\%  (0.5\%) of the 
\BzBzb\ (\BpBm) events. 

The selected sample of $\bre$ is used as normalization for the determination
of the branching fractions. 
It is contaminated by $\epem\ra q\bar q \ (q=u,d,s,c)$ events and by
other \FourS\to\BzBzb or \BpBm decays, 
in which the \bre\ is mistakenly reconstructed from particles 
coming from both $B$ mesons in the event.
To significantly reduce the $\epem\ra q\bar q$ background we require 
the angle $\theta_{TB}^*$, defined in the CM frame, between the thrust
axis~\cite{thrust} of the \bre\ and the thrust axis of all charged and neutral
particles in the event excluding the ones that form the \bre , to satisfy the requirement
$|\cos{\theta_{TB}^*}|<0.7$. 

\begin{figure}[htb!] 
 \begin{center} 
  \includegraphics[width=4.2cm]{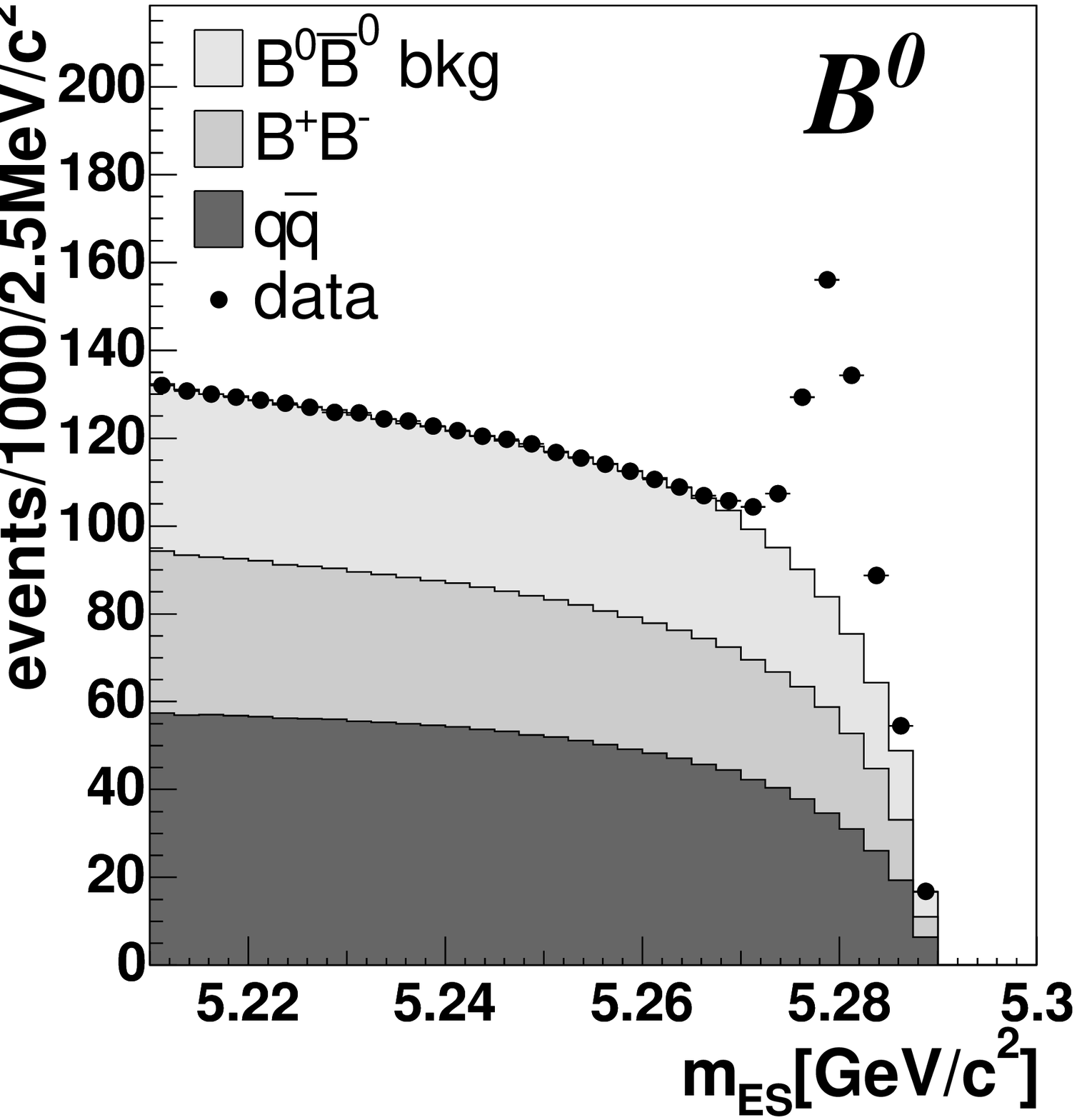} 
  \includegraphics[width=4.2cm]{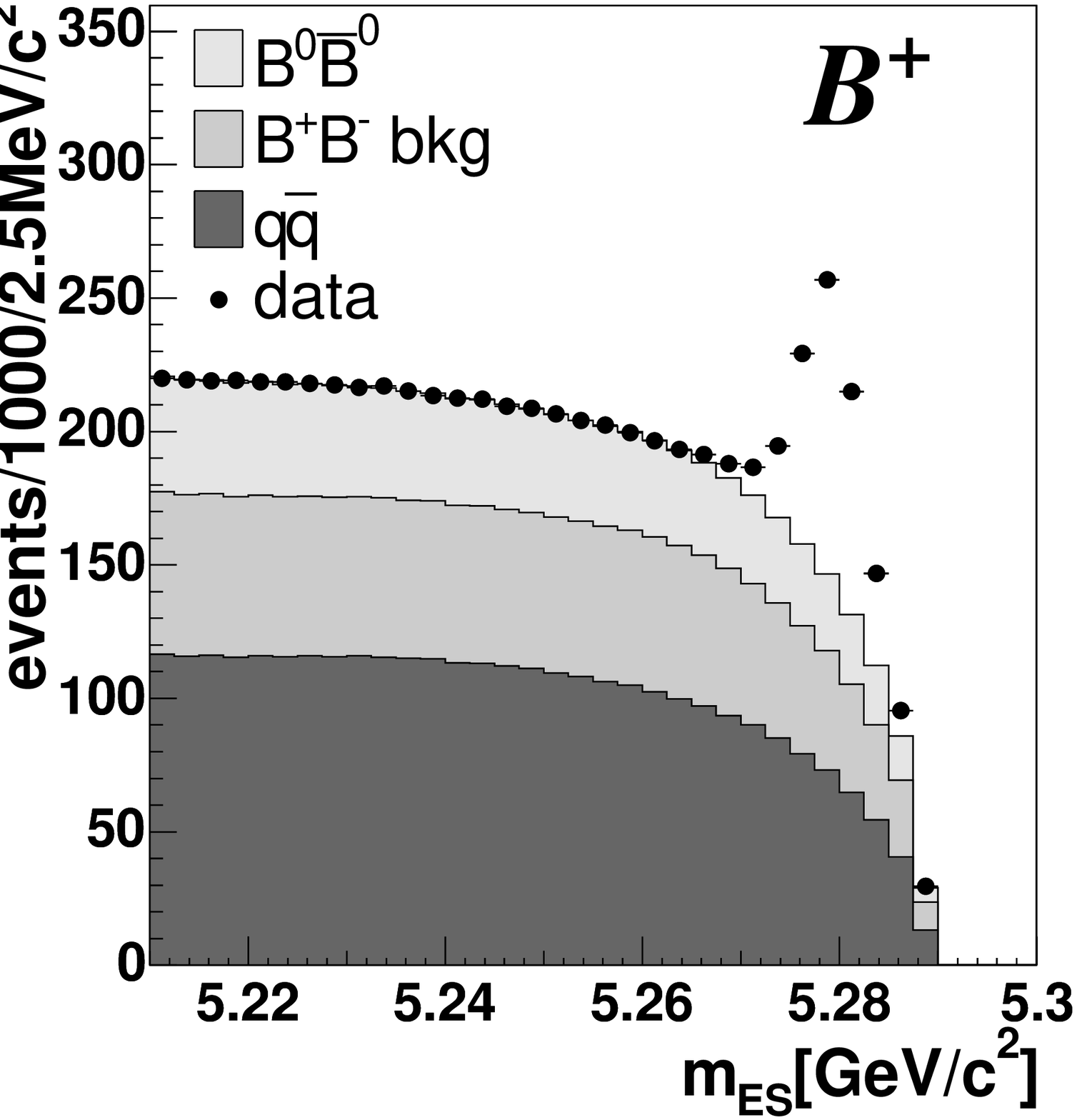} 
 \end{center} 
 
  \caption{ Distributions in \mes for the \bre\ sample. 
          The background contributions, determined as described in the text, are overlaid.  
} \label{fig:mesfit} 
\end{figure}

On this signal-enriched sample (Fig.~\ref{fig:mesfit}), the contributions from the 
background are estimated as the sum of three components: the 
$\epem\ra q\bar q$, the \BzBzb, and the \BpBm events. The shapes of 
these background distributions are taken from MC simulation.
The normalization of the $\epem\ra q\bar q$ background is taken from 
{\it off-resonance} data, scaled by the luminosity.
The normalization of the \BzBzb, \BpBm components are instead 
obtained by means of  a $\chi^2$ fit to the \mes distribution in 
the sideband region ($5.21\gevcc<\mes<5.26\gevcc$). 
The background contamination in the signal region ($\mes>5.27\gevcc$) is extrapolated and 
subtracted from the data to estimate the signal yield.
After correcting for the $|\cos{\theta_{TB}^*}|$ cut efficiency estimated in the MC,
the size of the total sample of fully reconstructed $B$ decays is 
$N_{\bre^{0}} = (2.90 \pm 0.01_{\rm stat.})\times 10^{5}$ and  
$N_{\bre^{+}} = (4.63 \pm 0.01_{\rm stat.}) \times 10^{5}$.

From the charged tracks and the neutral clusters that 
do not belong to the \bre\ we reconstruct the charmed mesons ($D_{\rm meas}$) in the modes $\Dz\to\Km\pip$,
 $\Km\pip\piz$, $\Km\pip\pim\pip$; $\Dp\to\Km\pip\pip$,  $\Dp\to\KS\pip$; and
$\Dsm\to\phi\pim$ ($\phi\to\Kp\Km$),  $\KS\Km$ ($\KS\to\pip 
\pim$), and $\Kstarz\Km$ ($\Kstarz\to\Kp\pim$).
We select $\phi$ and $\Kstarz$ candidates with a reconstructed mass 
within $15\mevcc$ and $70\mevcc$  from their nominal values~\cite{pdg2004}, respectively.
 The $D^*$ candidates are reconstructed in the decay modes $D^{*+}\to \Dz\pip$,
$\Dp\piz$, $\Dstarz\to \Dz\piz$, $\Dz\g$, and $\Dssm\to\Dsm\gamma$.
We require the reconstructed masses of the \Dz, \Dp , and \Dsm candidates and the  
 differences $\Delta m$ between the masses of the \Dstar and $D$ candidates 
to be within $1.5-3$ times its measured resolution from their nominal values~\cite{pdg2004},
depending on the background level.

We apply further selection criteria to enhance the signal contributions in the sample.
For $D^{(*)+}D_X^-$ and $D_{s}^{(*)-}D_X^{+}$ we consider 
neutral \bre\ candidates while for $D^{(*)0}D_X^-$ and $D_{s}^{(*)-}D_X^{0}$ we require
positive charged \bre\ candidates. 
We suppress background from $\B \to D^{(*)} l {\nu}$, while keeping events with a semileptonic
$D_X$ decay, by rejecting any event
with a remaining identified lepton with the appropriate charge and 
a momentum  in the \B rest frame ($p^*$) greater than $1 \gevc$.
In order to minimize the contamination of the modes with a $D^*$ to the modes
with a $D$ meson, we assign the events consistent with both the hypotheses 
($B\to DD_X$ and $B\to D^*D_X$) to the $D^*$ sample.

The invariant mass of $D_X$ (\mX ) is derived from the
missing four-momentum $p_{X} = p_{\Y4S}-p_{\bre} -p_{D_{\rm meas}}$, 
where all momenta are measured in the laboratory frame.
The \mX\ resolution is improved by a
global \FourS\  kinematic fit~\cite{Hulsbergen:2005pu} that includes beam position and energy
information and constrains the masses and decay vertices of the $D_{\rm meas}$. 
The $\chi^2$ of this fit is used to reduce the combinatorial background.
We remove reconstructed $D$ mesons with $\chi^2$ probability smaller than $0.1\%$.

Of the selected events, $3-6\%$ ($9-30\%$) contain multiple $D_{(s)}$ ($D_{(s)}^*$) candidates.
We retain those in the $D_{\rm meas}$ decay mode with the lowest combinatorial
background. If there are multiple candidates with the same
decay mode, we select the one with the lowest value of $|m_D-m_{PDG}|$ and
$(m_{D_{\rm meas}}-m_{PDG})^2/\sigma_{m_{D_{\rm meas}}}^2 + (\Delta m-\Delta m_{PDG})^2/\sigma_{\Delta m}^2$
for $D_{(s)}$ and $D_{(s)}^*$ respectively, where
$m$ is the reconstructed mass of the $D_{\rm meas}$ candidate and the subscript $PDG$ indicates nominal values~\cite{pdg2004}.

Finally, we consider only candidates in the range 
$1.65\gevcc<\mX<2.71\gevcc$ for the $D^{(*)+/0}D_X$  modes and
$1.68\gevcc<\mX<2.31\gevcc$ for $D_s^{(*)-} D_X$. These ranges were chosen to 
minimize the total uncertainty  introduced by the background shape and normalization.

\begin{figure*}[!] 
\begin{center} 
 \includegraphics[width=4.2 cm]{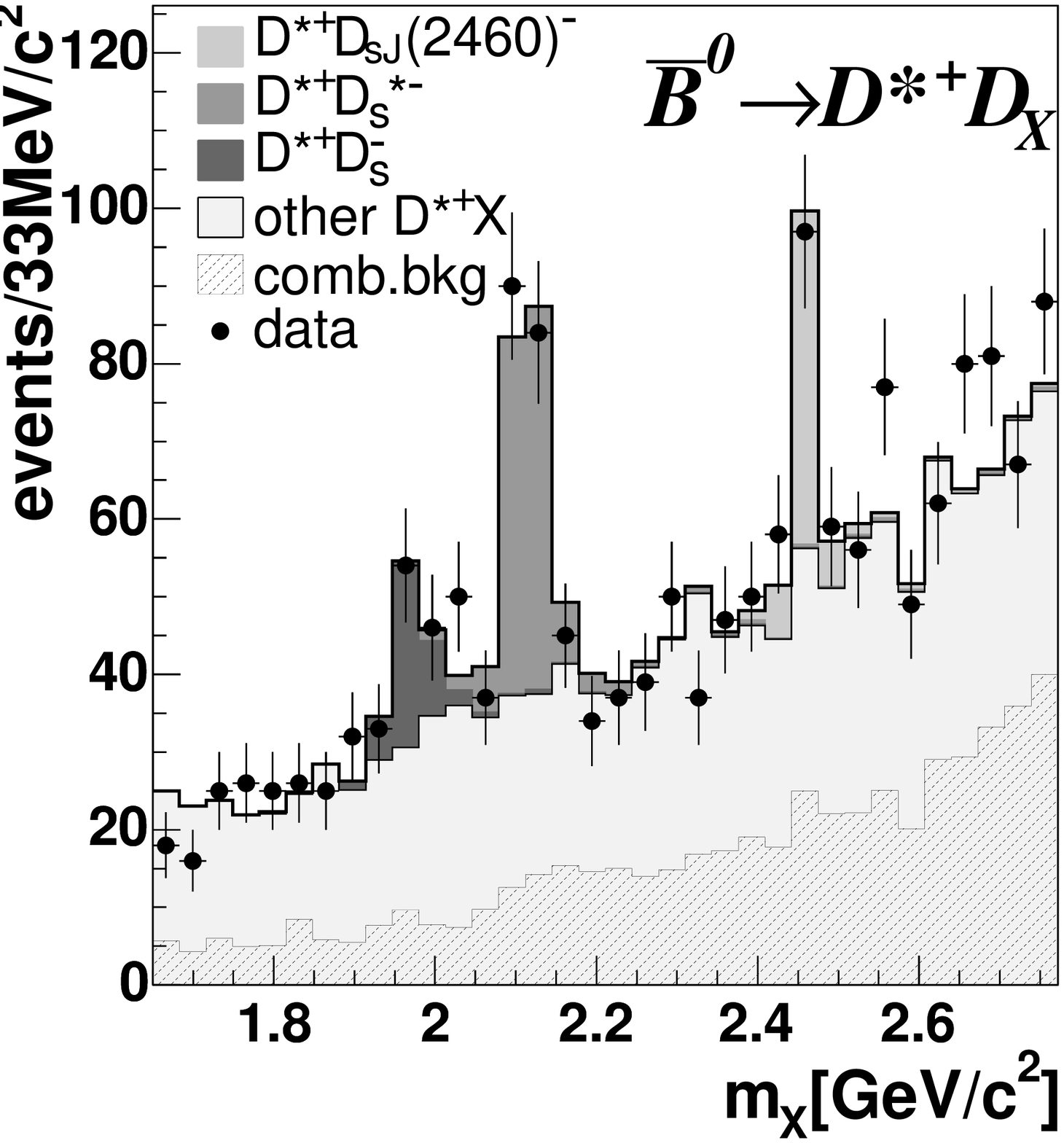} 
 \includegraphics[width=4.2 cm]{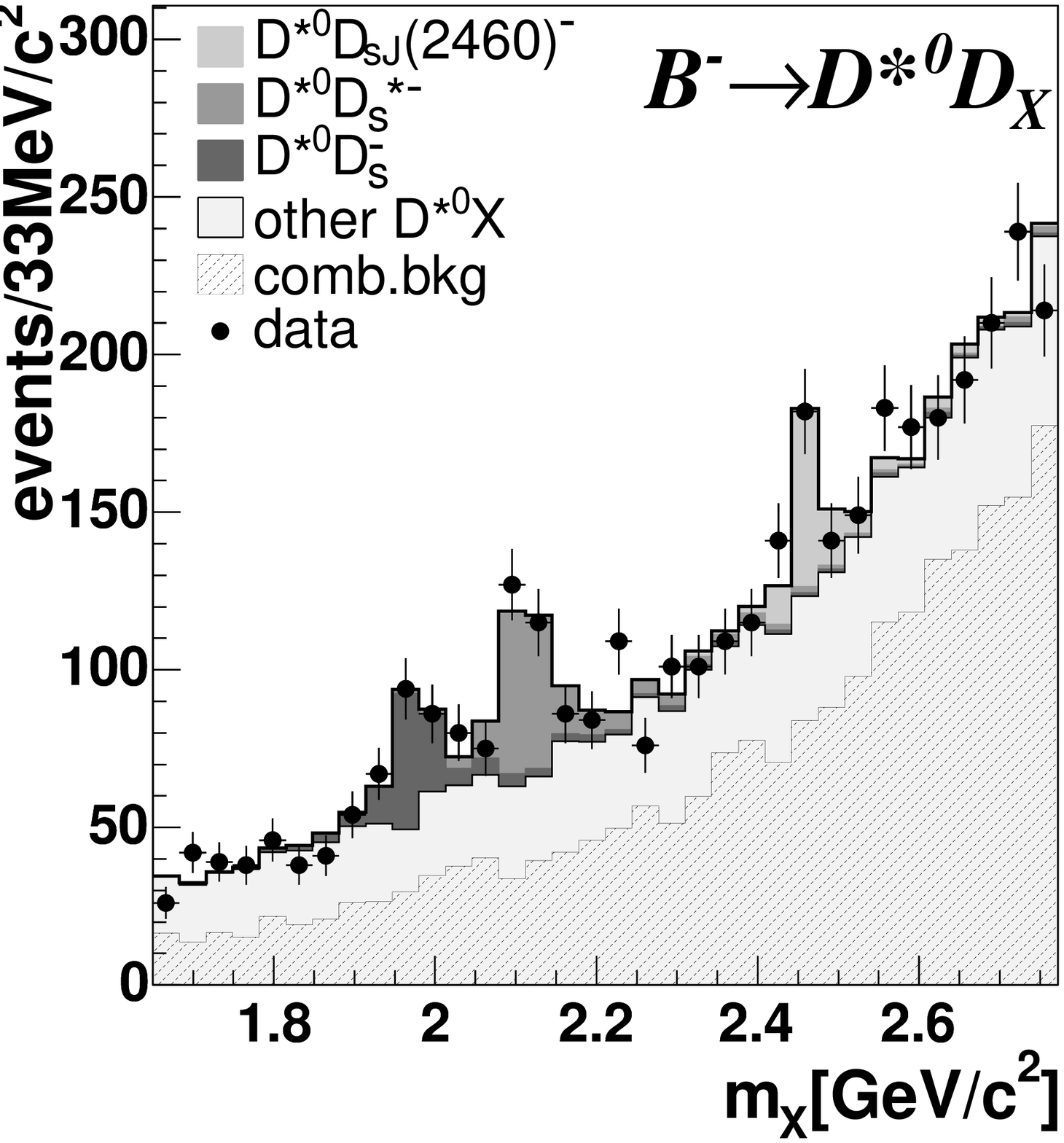} 
 \includegraphics[width=4.2 cm]{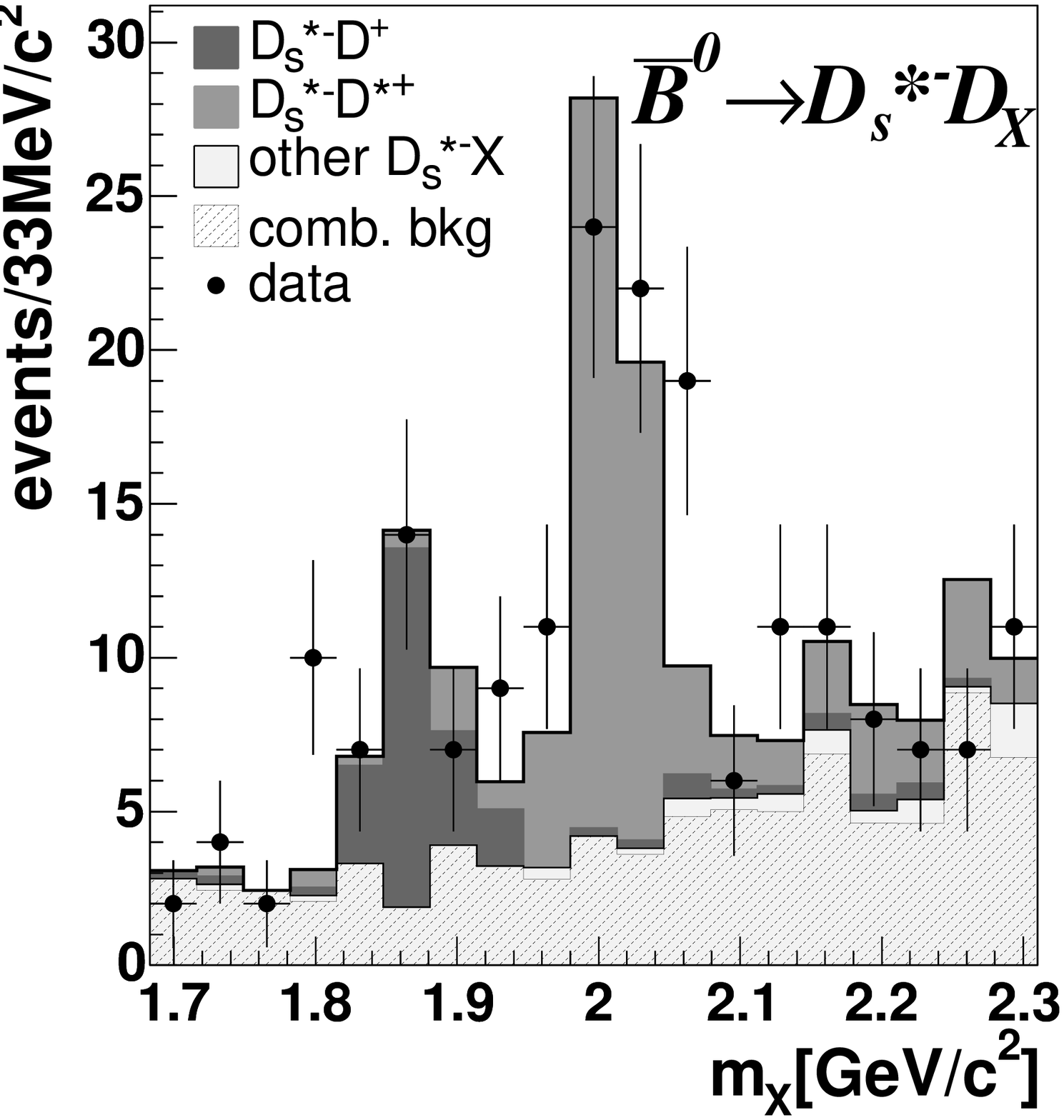} 
 \includegraphics[width=4.2 cm]{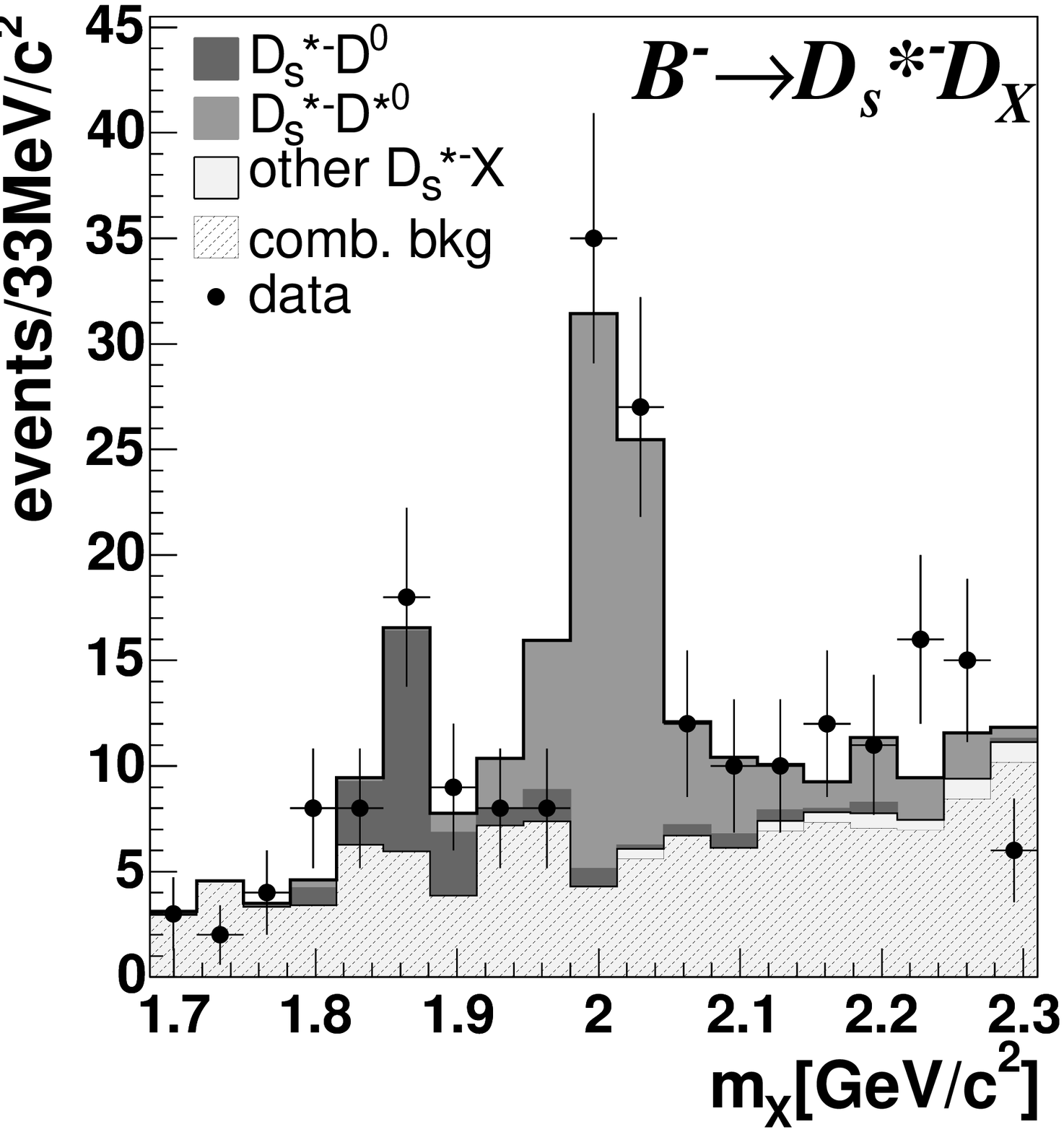}\\ 
 \includegraphics[width=4.2 cm]{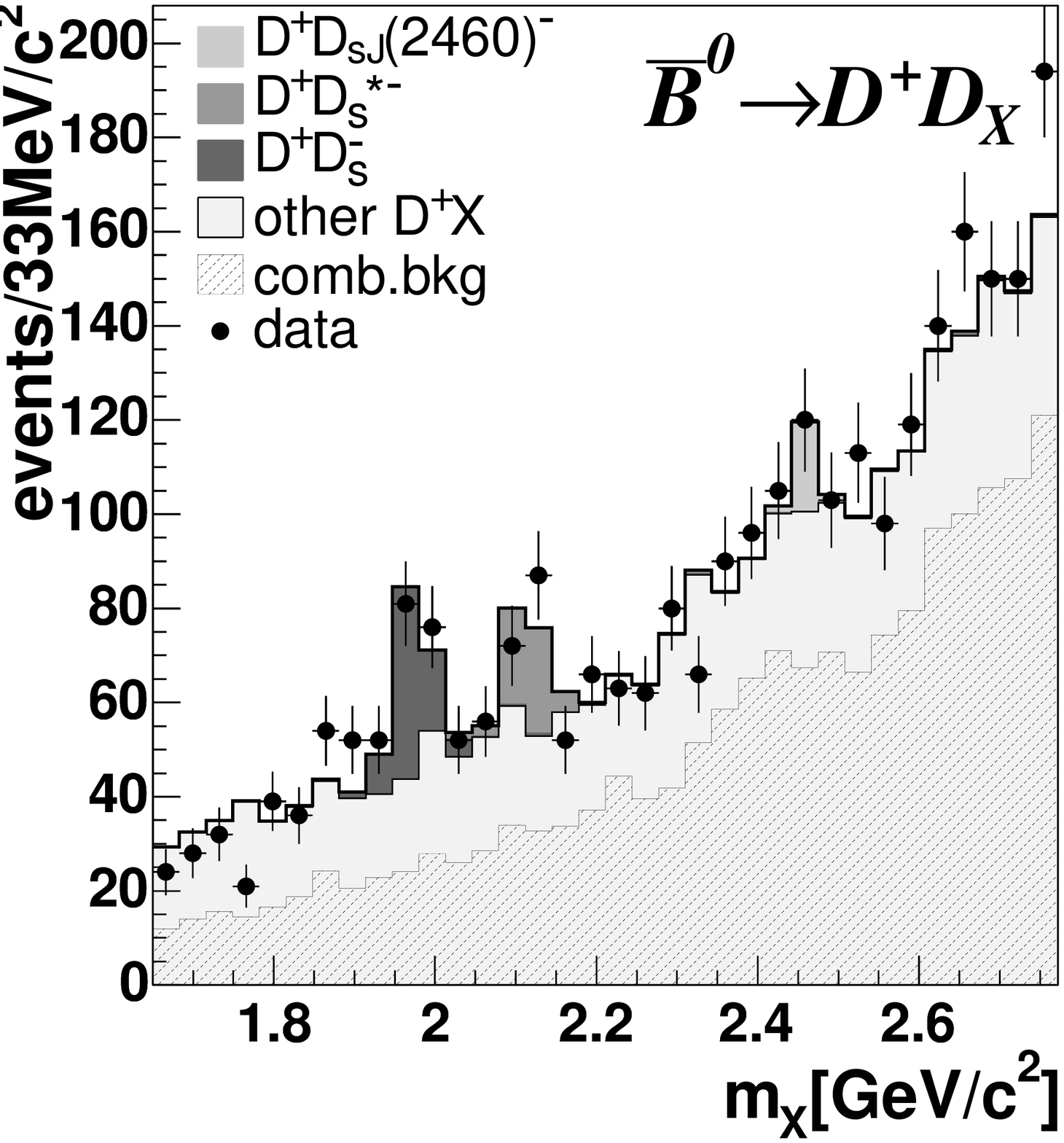}  
 \includegraphics[width=4.2 cm]{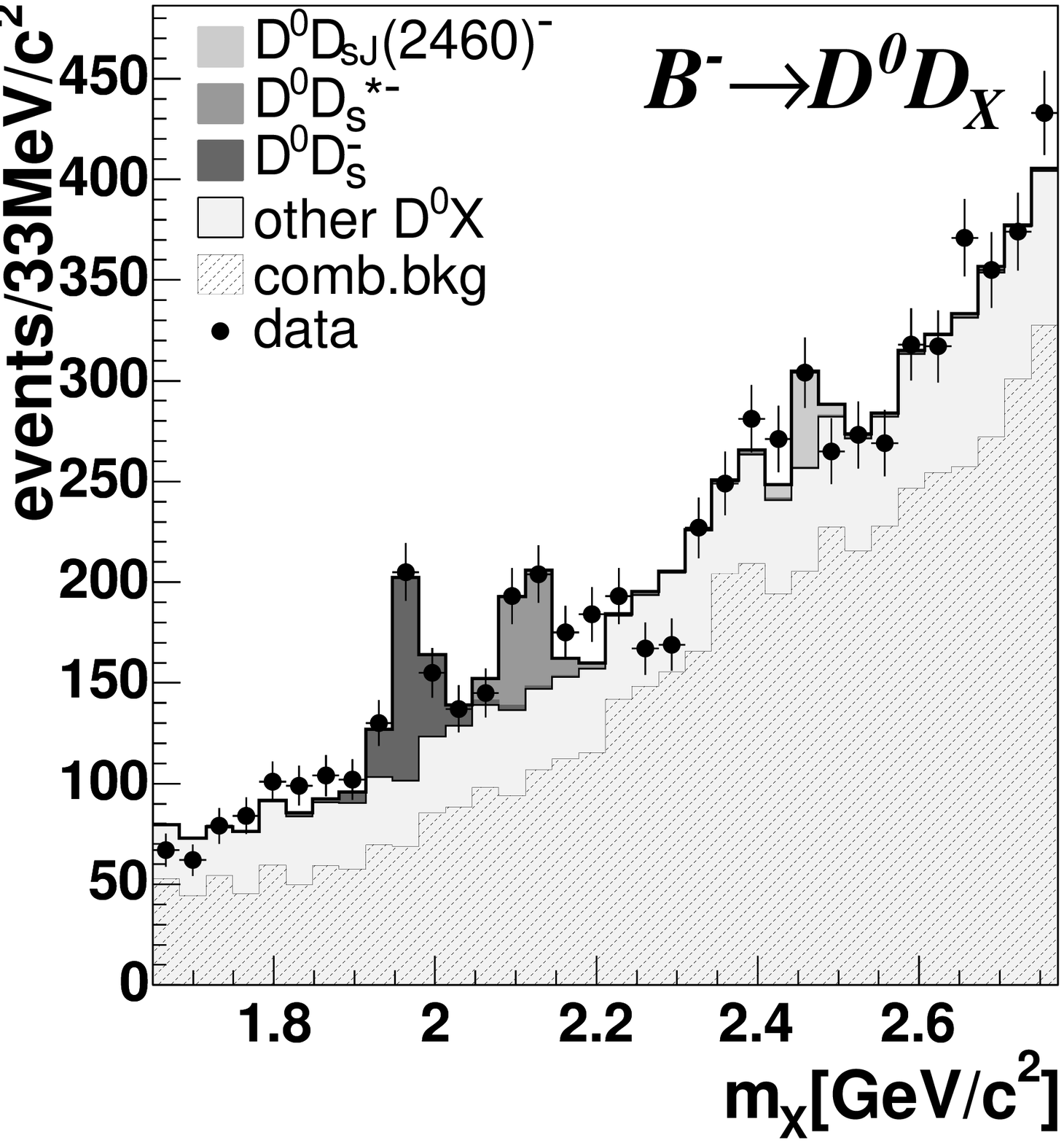} 
 \includegraphics[width=4.2 cm]{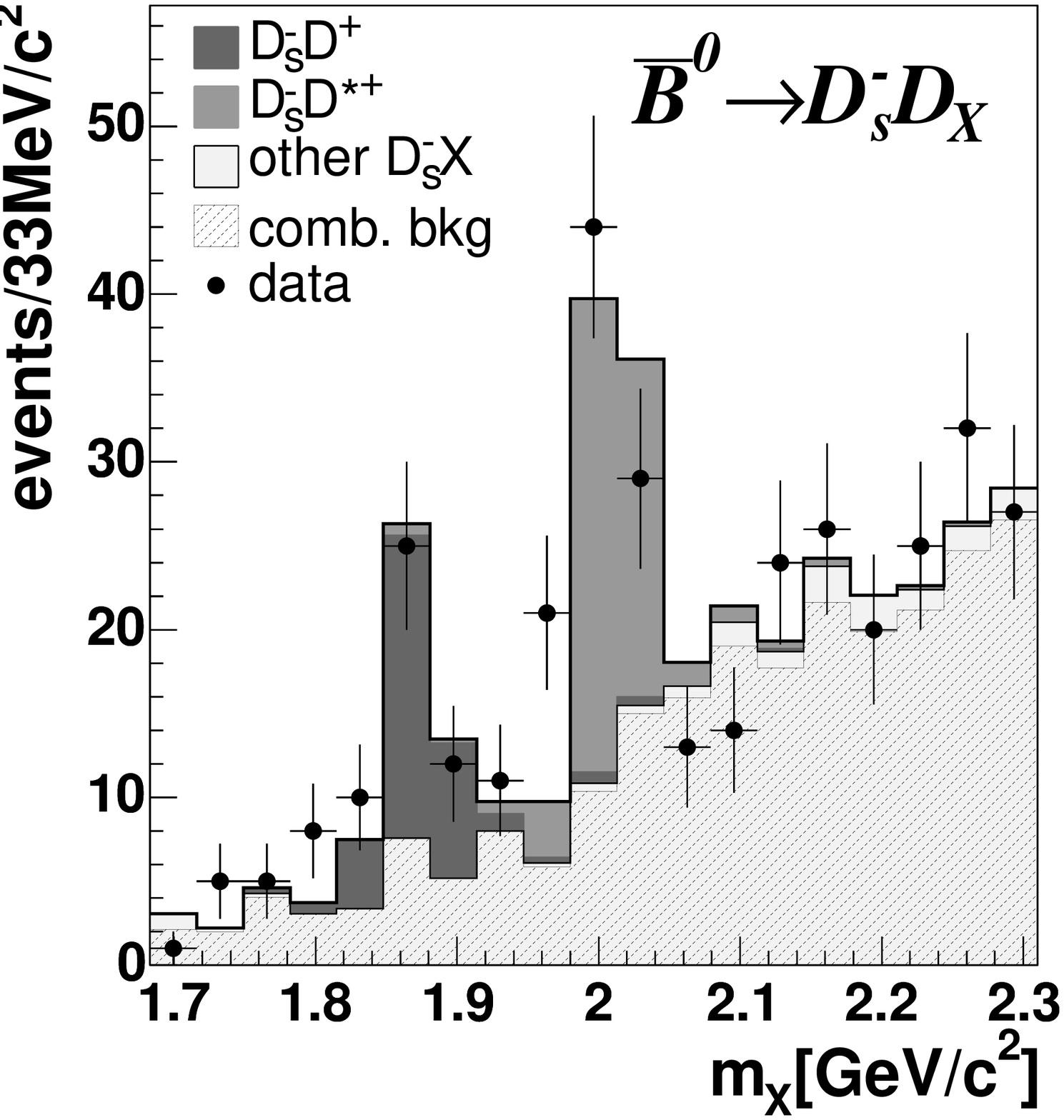}  
 \includegraphics[width=4.2 cm]{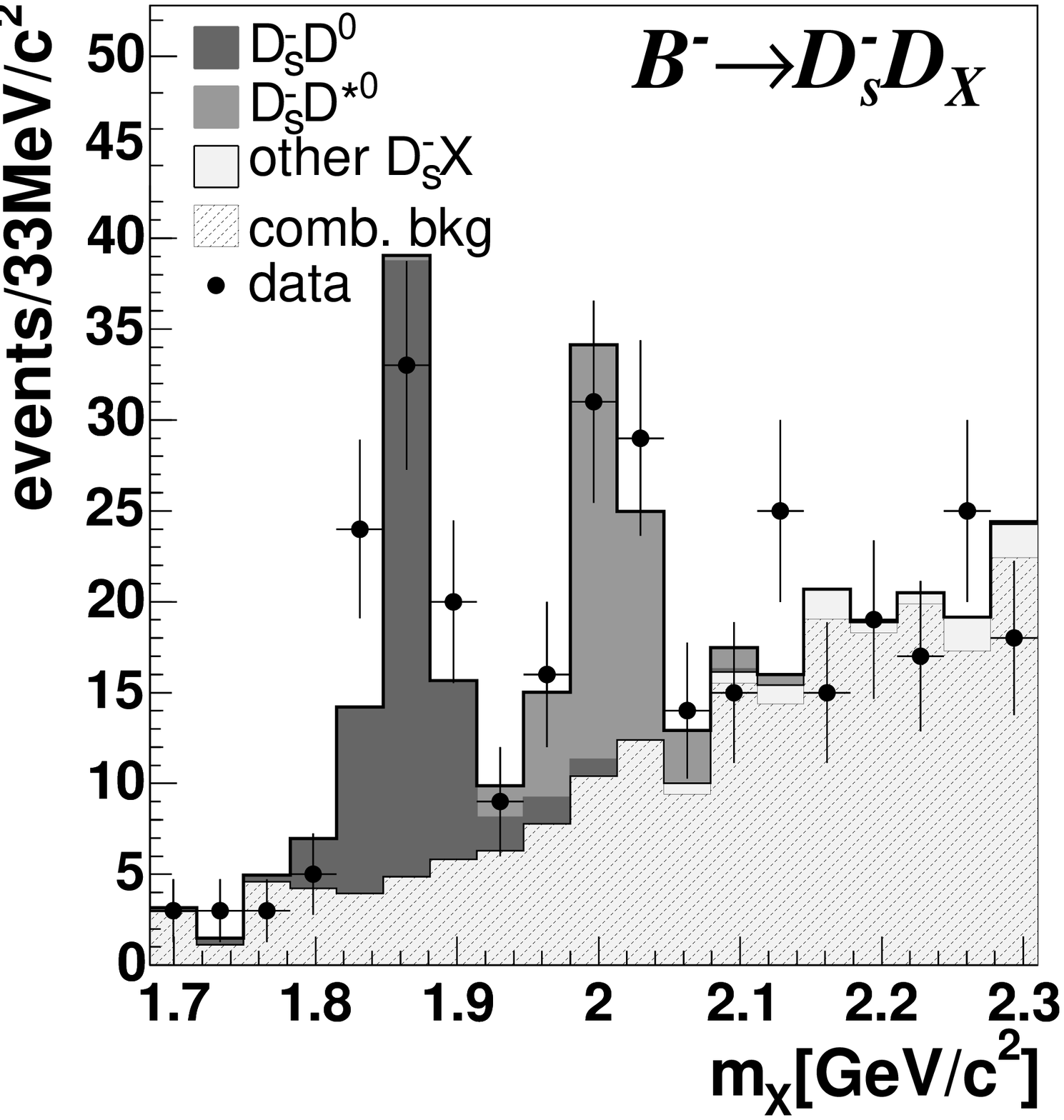} 
 \end{center} 
\caption{ Distributions of \mX. Fitted $\Bb \ra D^{(*)+,0}D_s^{(*)-}$ and 
$\Bb \ra D^{(*)+,0}\Dsjsm$ signal contributions and background components,  
determined as described in the text, are overlaid to the data points. 
}\label{fig:results} 
\end{figure*}

The yield of each decay mode is extracted from the \mX\ distribution by a binned $\chi^2$ fit 
of a sum of $n_{\rm sig}$ signal contributions ($N^{\rm sig}$) and
the total background contribution  ($N^{\rm bkg}$), which is a sum of 
the combinatorial background, other $B \ra D_{(s)}^{(*)}D_X$ decays, and $D_{(s)}^{(*)}-D_{(s)}$
crossfeed, to the experimental data. The signal and background distributions are histograms taken from MC simulation. 
For $D^{0}D_X$ we also weight the background shape with a second order polynomial function
whose parameters are fitted on data.
In the case of 
 $D^{(*)+/0}D_X$ modes we consider three signal components: $D^{(*)+/0} \Dsm$, $D^{(*)+/0} \Dssm$, and $D^{(*)+/0} \Dsjsm$,
while in the case of  $D_s^{(*)-} D_X$ modes we consider two signal components: $D_s^{(*)-}D^{+/0}$ and $D_s^{(*)-}D^{*+/0}$. 
The $\chi^2$ is defined as:
\begin{equation}
\label{eq:chi2fit}
\nonumber 
\chi^2(C_j,C_{\rm bkg})=
\sum_i \left( \frac{N_i^{\rm meas}  - \mu_i(C_j,C_{\rm bkg})}{\sqrt{\delta {N_i^{\rm meas}}^2+\delta {N_i^{MC}}^2}} \right)^2
\end{equation}
where 
$N_i^{\rm meas}$ is the number of observed events in bin $i$, $\mu_i$ corresponds to
$\mu_i= \sum_{j=1,n_{sig}} C_j N^{\rm sig}_{ij} + C_{\rm bkg} N^{\rm bkg}_i$, the index $j$ denotes
the signal component, and $\delta {N_i^{\rm meas}}$ and $\delta {N_i^{MC}}$ 
are the statistical uncertainties for data and MC samples, respectively.
The relative normalizations of each component ($C_j$ and $C_{\rm bkg}$) are allowed to vary in the fit.
The measured \mX\ distributions and the results of the fits are shown in Fig.~\ref{fig:results}.   

The branching fractions are extracted as $\BR (f) = N_{\rm fit}/(\eps N_{\bre})$, 
where $N_{\rm fit}$ is the number of signal events obtained from the fit to the 
\mX\ distribution for a given mode 
and  $\eps$, which includes the intermediate branching fractions of $D_{\rm meas}$ and 
its decay products, is the selection efficiency estimated using MC simulation.

The dominant systematic uncertainties originate from the lack of knowledge of the correct shapes 
used in the \mX\ fit, and from the determination of efficiencies (because of  
the limited MC statistics). These uncertainties range from $5.6\%$ to $25\%$, depending on the mode.
The systematic uncertainties due to the determination of $N_{\bre}$ and 
to the differences between data and MC in the composition of
the reconstructed \bre\ modes range between $3.7\%$ and $6.7\%$ for \Bz, and between  
$3.5\%$ and $9.0\%$ for \Bp\, depending on the mode under study.  
Other uncertainties come from track reconstruction efficiency  
(1.4\% per track and 2.2\% per soft pion), $\gamma$ and 
\piz efficiencies ($3.0\%$ per \piz and 1.8\% per  
$\gamma$), and kaon identification ($2\%$ per kaon).
The uncertainties due to branching fraction measurements for exclusive $D_{(s)}^{(*)}$
decays ~\cite{pdg2004} contribute between 3.0\% and 7.4\%, depending on the mode. 
We check the uncertainties introduced by the $\chi^2$ cut of the kinematic fit by comparing data and MC 
control samples for $\B \to D^{(*)} l {\nu}$ obtained with
all previously mentioned cuts except for the $p^* > 1 \gevc$ criterion applied. 
The statistical uncertainty of this comparison is
used as the systematic uncertainty (between 0.5\% and 2.3\%). 

We combine the sixteen measurements of $\B \ra D^{(*)}D_{s(J)}^{(*)}$ to
obtain the eight branching fractions for these modes and $\BR(\Dsm \ra \phi \pim)$ 
in a $\chi^2$ fit.  
In this combination the ratios $\BR(\Dsm\ra \Kstarz\Km)/\BR(\Dsm \ra \phi \pim)$ 
and $\BR(\Dsm \ra \KS\Km)/\BR(\Dsm \ra \phi \pim)$,
included in the efficiency calculation when $D_{\rm meas}=D^{(*)-}_s$, are fixed
\cite{pdg2004}, while $\BR(\Dsm \ra \phi \pim)$ is a free parameter.
The MC model used to generate the $\Dsm \ra \Kp\Km\pim$ decays
does not include any interference among the different final
states ($\phi \pim$, $\Kstarz\Km$, $f_0(980)\pim$, ...).
Correlated and uncorrelated uncertainties are properly taken into account in the covariance matrix.  
The results of this fit are given in the last column of Table~\ref{tab:results}.
\begin{table*}[]
  \centering
  \caption{Event yields ($N_{\rm fit}$), efficiencies ($\eps$), and branching fractions (\BR)  for pairs of detected decay modes, separately and
combined. In this combination we use only the results in this paper. $\BR(\Dsm \ra \phi \pim)$ is a free parameter and is also 
reported in the table. The first uncertainty on ${\cal B}$ is statistical, the second is systematic.  The parameter $\it{k}$ corresponds to 
$\it{k} =$ 3.6\%$/(\BR(\Dsm \ra \phi\pim))$.
} \label{tab:results}
\begin{tabular}{l|l|r@{$\,\pm\,$}r@{    }lcc|c}
  \hline\hline
  Decay mode & $D_{\rm meas}$ 
  & \multicolumn{3}{c}{$N_{\rm fit}$} & $\eps (\%)$ & ${\cal B}(\%)$  & Combined ${\cal B}(\%)$ \\
  \hline\hline
\rule[0mm]{0mm}{1.1em}\multirow{2}{*}{$\Bzb\ra D_{s}^-D^{+}$}     & $D^{+}$ & $86$&17&   & $3.29\pm0.16$ & $0.90\pm 0.18\pm 0.14 $& \multirow{2}{*}{$0.64 \pm 0.13\pm 0.10 $} \\ 
\rule[0mm]{0mm}{1.1em}                                           & $D_{s}^-$ & $39$&9&   & $1.79\pm0.12$ & $(0.74\pm 0.17\pm 0.13) \cdot \it{k}$ \\  
\rule[0mm]{0mm}{1.1em}\multirow{2}{*}{$\Bzb\ra D_{s}^{*-}D^{+}$}  & $D^{+}$ & $63$&19&   & $3.24\pm0.16$ & $0.67\pm 0.20\pm 0.11 $& \multirow{2}{*}{$0.69 \pm 0.16\pm 0.09 $} \\  
\rule[0mm]{0mm}{1.1em}                                           & $D_s^{*-}$ & $30$&9&  & $0.91\pm0.08$ & $(1.15\pm 0.33\pm 0.26) \cdot \it{k} $& \\   
\rule[0mm]{0mm}{1.1em}\multirow{2}{*}{$\Bzb\ra D_{s}^-D^{*+}$}    & $D^{*+}$ & $48$&13&  & $2.86\pm0.13$ & $0.57\pm 0.16\pm 0.09 $& \multirow{2}{*}{$0.71 \pm 0.13\pm 0.09 $} \\  
\rule[0mm]{0mm}{1.1em}                                           & $D_s^{-}$ & $68$&12&  & $1.63\pm0.10$ & $(1.42\pm 0.26\pm 0.20) \cdot \it{k} $& \\   
\rule[0mm]{0mm}{1.1em}\multirow{2}{*}{$\Bzb\ra D_{s}^{*-}D^{*+}$} & $D^{*+}$ & $129$&18& & $2.68\pm0.09$ & $1.65\pm 0.23\pm 0.19 $& \multirow{2}{*}{$1.68 \pm 0.21\pm 0.19 $} \\   
\rule[0mm]{0mm}{1.1em}                                           & $D_s^{*-}$ & $84$&14& & $0.86\pm0.05$ & $(3.38\pm 0.60\pm 0.61) \cdot \it{k} $& \\    
\rule[0mm]{0mm}{1.1em}\multirow{2}{*}{$B^-\ra D_{s}^-D^{0}$}     & $D^{0}$ & $214$&28&   & $3.46\pm0.11$ & $1.33\pm 0.18\pm 0.32 $& \multirow{2}{*}{$0.92 \pm 0.14\pm 0.18 $} \\  
\rule[0mm]{0mm}{1.1em}                                           & $D_s^{-}$ & $66$&10&  & $1.28\pm0.07$ & $(1.11\pm 0.17\pm 0.17) \cdot \it{k} $& \\   
\rule[0mm]{0mm}{1.1em}\multirow{2}{*}{$B^-\ra D_{s}^{*-}D^{0}$ } & $D^{0}$ & $160$&31&   & $3.71\pm0.12$ & $0.93\pm 0.18\pm 0.19 $& \multirow{2}{*}{$0.77 \pm 0.15\pm 0.13 $} \\   
\rule[0mm]{0mm}{1.1em}                                           & $D_s^{*-}$ & $26$&10& & $0.64\pm0.05$ & $(0.87\pm 0.33\pm 0.16) \cdot \it{k} $& \\    
\rule[0mm]{0mm}{1.1em}\multirow{2}{*}{$B^-\ra D_{s}^-D^{*0}$}    & $D^{*0}$ & $152$&29&  & $2.69\pm0.10$ & $1.21\pm 0.23\pm 0.20 $& \multirow{2}{*}{$0.76 \pm 0.15\pm 0.13 $} \\   
\rule[0mm]{0mm}{1.1em}                                           & $D_s^{-}$ & $52$&11&  & $1.33\pm0.07$ & $(0.82\pm 0.18\pm 0.10) \cdot \it{k} $& \\    
\rule[0mm]{0mm}{1.1em}\multirow{2}{*}{$B^-\ra D_{s}^{*-}D^{*0}$} & $D^{*0}$ & $216$&33&  & $2.73\pm0.07$ & $1.70\pm 0.26\pm 0.24 $& \multirow{2}{*}{$1.62 \pm 0.22\pm 0.18 $} \\    
\rule[0mm]{0mm}{1.1em}                                           & $D_{s}^{*-}$&$90$&15& & $0.82\pm0.04$ & $(2.38\pm 0.41\pm 0.31) \cdot \it{k} $& \\     
\hline
\rule[0mm]{0mm}{1.1em}$\Dsm   \ra \phi\pim$ & - & \multicolumn{3}{c}{-} & - & - &$4.58\pm 0.48\pm 0.68 $ \\    
\hline
\rule[0mm]{0mm}{1.1em}$\Bzb\ra \Dsjsm D^{+}$ & $D^{+}$ &   $27$&16& & $3.61\pm0.27$ & $0.26\pm 0.15\pm 0.07 $& \\     
\rule[0mm]{0mm}{1.1em}$\Bzb\ra \Dsjsm D^{*+}$ & $D^{*+}$ & $64$&15& & $2.51\pm0.15$ & $0.88\pm 0.20\pm 0.14 $& \\      
\rule[0mm]{0mm}{1.1em}$B^-\ra \Dsjsm D^{0}$ & $D^{0}$ &    $75$&28& & $3.78\pm0.24$ & $0.43\pm 0.16\pm 0.13 $& \\      
\rule[0mm]{0mm}{1.1em}$B^-\ra \Dsjsm D^{*0}$ & $D^{*0}$  & $147$&34& & $2.81\pm0.14$ & $1.12\pm 0.26\pm 0.20 $& \\       
   \hline\hline
\end{tabular}
\end{table*}

We further combine the results of this analysis with $\Bb \ra D^{(*)+/0}D_s^{(*)-}$ 
exclusive branching fractions from ~\cite{Gibaut:1995tu,Ahmed:2000ad,nodsphipi,Aubert:2003jj} and 
the \babar\ results for $\BR(\Bb \ra \Dsjsm D^{(*)})$ \cite{Aubert:2004pw}, obtaining 
the following branching fractions: 
\begin{eqnarray}
\nonumber 
\BR(\Dsjsm \ra \Dssm \piz) = (56 \pm 13_{\rm stat.} \pm 9_{\rm syst.})\% , \\ 
\nonumber  
 \BR(\Dsjsm \ra \Dsm \gamma)  = (16 \pm 4_{\rm stat.} \pm 3_{\rm syst.})\%, \\  
\nonumber  
 \BR(\Dsm   \ra \phi\pim)  = (4.62 \pm 0.36_{\rm stat.} \pm 0.50_{\rm syst.})\% .  
\nonumber  
\end{eqnarray} 
 
In conclusion, we have measured the branching fractions for the decays $\Bb \ra D^{(*)+,0}D_s^{(*)-}$. 
These are consistent with the existing measurements~\cite{pdg2004} and, in several cases, have a  
significantly smaller uncertainty. The combination of these results with the existing measurements  
provide the branching fraction for $\Dsm \ra \phi\pim$, which is also consistent with  
the most recent measurement \cite{Aubert:2005xu} and confirms a larger value
compared to the previous world average~\cite{pdg2004}. 
We have extracted the absolute branching fractions for  
$\Bb \ra D^{(*)+,0}\Dsjsm$, thus allowing the first measurement  
of the \Dsjsm\ decay rates.  
Our results show that the $\Dsjsm$ meson decays via photon or \piz emission to $D_s^{(*)-}$
in $(72\pm 19)\%$ of the cases.

We are grateful for the excellent luminosity and machine conditions
provided by our \pep2\ colleagues, 
and for the substantial dedicated effort from
the computing organizations that support \babar.
The collaborating institutions wish to thank 
SLAC for its support and kind hospitality. 
This work is supported by
DOE
and NSF (USA),
NSERC (Canada),
IHEP (China),
CEA and
CNRS-IN2P3
(France),
BMBF and DFG
(Germany),
INFN (Italy),
FOM (The Netherlands),
NFR (Norway),
MIST (Russia), and
PPARC (United Kingdom). 
Individuals have received support from the
Marie Curie EIF (European Union) and
the A.~P.~Sloan Foundation.


\begin{thebibliography}{99}
\bibitem{charge-conj} 
Charge conjugate reactions are implied throughout this paper.
\end{thebibliography}
\end{document}